\documentclass[aps,pra,longbibliography,showkeys,showpacs,groupedaddress,twocolumn,10pt]{revtex4-1}
\usepackage{amssymb}
\usepackage{graphicx}
\usepackage{dcolumn}
\usepackage{bm}
\usepackage{amsmath}
\usepackage{subfigure}
\usepackage{float}
\usepackage{color}

\begin{document}
\title{Atom-based radio-frequency field calibration and polarization measurement using cesium $nD_J$ Floquet states}

\author{Yuechun Jiao$^{1,3}$}
\author{Liping Hao$^{1,3}$}
\author{Xiaoxuan Han$^{1,3}$}
\author{Suying Bai$^{1,3}$}
\author{Georg Raithel$^{1,2}$}
\author{Jianming Zhao$^{1,3}$}
\thanks{Corresponding author: zhaojm@sxu.edu.cn, graithel@umich.edu}
\author{Suotang Jia$^{1,3}$}

\affiliation{$^{1}$State Key Laboratory of Quantum Optics and Quantum Optics Devices, Institute of Laser Spectroscopy, Shanxi University, Taiyuan 030006, China}
\affiliation{$^{2}$ Department of Physics, University of Michigan, Ann Arbor, Michigan 48109-1120, USA}
\affiliation{$^{3}$Collaborative Innovation Center of Extreme Optics, Shanxi University, Taiyuan 030006, China}

\begin{abstract}
We investigate atom-based electric-field calibration and polarization measurement of a 100-MHz linearly polarized radio-frequency (RF) field using cesium Rydberg-atom electromagnetically induced transparency (EIT) in a room-temperature vapor cell.
The calibration method is based on matching experimental data with the results of a theoretical Floquet model. The utilized 60$D_J$ fine structure Floquet levels exhibit $J$- and $m_j$-dependent AC Stark shifts and splittings, and develop even-order RF-modulation sidebands. The Floquet map of cesium 60$D_J$ fine structure states exhibits a series of exact crossings between states of different $m_j$, which are not RF-coupled. These exact level crossings are employed to perform a rapid and precise ($\pm 0.5\%$) calibration of the RF electric field.
We also map out three series of narrow avoided crossings between fine structure Floquet levels of equal $m_j$ and different $J$, which are
weakly coupled by the RF field via a Raman process. The coupling
leads to narrow avoided crossings that can also be applied as spectroscopic markers for RF field calibration.
We further find that the line-strength ratio of intersecting Floquet levels with different $m_j$ provides
a fast and robust measurement of the RF field's polarization.
\end{abstract}
\keywords{Rydberg EIT, atom-based field calibration, polarization and amplitude of RF field}
\pacs{32.80.Rm, 42.50.Gy, 32.30.Bv}
\maketitle

\section{Introduction}

In the past decades atom-based metrology has had an enormous impact on science, technology and everyday life. Seminal advances include microwave and optical atomic clocks~\cite{Heavner2014, Ludlow2015}, the global positioning system, and highly sensitive, position-resolved magnetometers~\cite{Savukov2005, Patton2012}. Atom-based field measurement has clear advantages over other field measurement methods because it is calibrating-free, due to the invariance of atomic properties. Atom-based metrology has recently expanded into electric-field measurement. An all-optical sensing approach employed by numerous groups is electromagnetically induced transparency (EIT) of atomic vapors, utilizing Rydberg levels~\cite{Mohapatra2007} to measure the properties of the electric field. Rydberg atoms are well-suited for this purpose owing to their extreme sensitivities to DC and AC electric fields, which manifest in large DC polarizabilities and microwave-transition dipole moments~\cite{Gallagher1994}. Developments include measurements of microwave fields and polarizations~\cite{Sedlacek2012, Sedlacek2013, Fan2015}, millimeter waves~\cite{Gordon2014}, static electric fields~\cite{Barredo2013}, and sub-wavelength imaging of microwave electric-field distributions~\cite{Fan2014, Holloway2014a}. In the frequency range from 10's to 100's of MHz, Rydberg-EIT RF modulation spectroscopy is a promising method to accomplish atom-based, calibration-free RF electric-field measurement~\cite{Bason2010,Jiao2016}. Rydberg-EIT in vapor cells offers significant potential for miniaturization~\cite{Budker2007,Daschner2014} of the RF sensor.

Accurate calibration of the electric field is important, for instance, for antenna calibration, characterization of electronic components, etc. Conventional calibration with field sensors that involve dipole antennas (that need to be calibrated first) obviously leads into a chicken-and-egg dilemma~\cite{Holloway2014}. In the present work we provide an atom-based calibration method for vector electrometry of RF-fields using Rydberg-EIT in cesium vapor cells. The basic idea is that the RF generates a series of intersections between levels in the Rydberg Floquet map (a map in which field-perturbed Floquet level energies are plotted versus RF electric field).
The (anti-)crossings occur between Floquet states originating in fine-structure components of $nL_J$ states with equal principal and angular-momentum quantum numbers $n$ and $L$. The crossings present excellent field markers that we use as calibration points for the electric-field strength.
Specifically, we measure the RF-dressed Cs~60$D_J$ states via Rydberg-EIT spectroscopy at a test frequency of 100~MHz. The dependence of the Floquet spectrum on the strength and the polarization of the RF field is investigated.
There are exact crossings between states of different $m_j=1/2$, $3/2$ and $5/2$, which are not coupled in a linearly polarized RF field.
The crossings provide field markers, which we use to calibrate the field strength in a test RF transmission system.
We also analyze narrow, spectroscopically resolved anti-crossings between Floquet states of equal $m_j=1/2$ (or $3/2$), and different $J=3/2$ and $5/2$. Transitions between those states are allowed via an RF Raman process.
Further, the EIT line-strength ratios of intersecting Floquet states with unlike $m_j$ yield the field polarization. At various stages of the work, the measured spectroscopic data are matched with the results of Floquet calculations to accomplish the calibration tasks.

\section{Experimental Setup}

\begin{figure}[ht]
\centering
\includegraphics[width=0.45\textwidth]{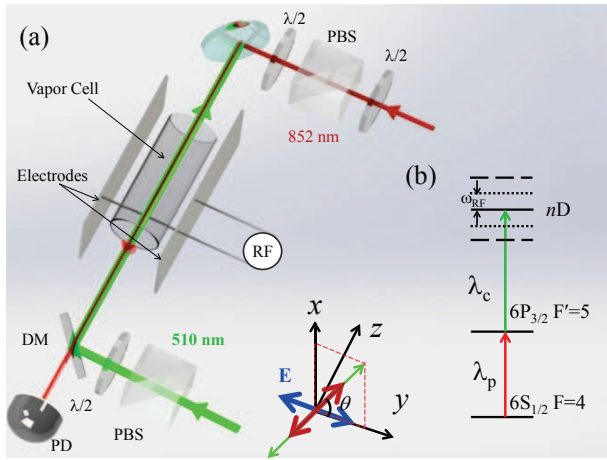}
\caption{(a) Schematic of the experimental setup. The coupling and probe beams counter-propagate through a cesium vapor cell along the $\hat{z}$-axis. Two electrode plates are located on the both sides of the vapor cell, where the RF field is applied. The polarization of the RF field $\bf{E}$ (blue arrow) points along the $\hat{y}$-axis. The polarization of the beams (red and green arrows) can be adjusted with the $\lambda$/2 wave plates to form an angle $\theta$ with the RF electric field $\bf{E}$. The probe beam is passed through a dichroic mirror (DM) and is detected with a photodiode (PD).
Polarizing beam splitters (PBS) are used to produce beams with pure linear polarizations. (b) Energy level scheme of cesium Rydberg-EIT transitions. The probe laser, $\lambda_{p}$, is resonant with the lower transition, $|6 S_{1/2}, F=4\rangle $ $\to$ $|6P_{3/2}, F'=5\rangle $, and the coupling laser, $\lambda_{c}$, is scanned through the Rydberg transitions $|6P_{3/2}, F'=5 \rangle $ $\to$ $|60D_J\rangle$. The applied RF electric field (frequency $\omega_{RF}$ = 2$\pi \times$100~MHz) produces AC Stark shifts and RF modulation sidebands that are separated in energy by even multiples of $\hbar \omega_{RF}$.}
\end{figure}

A schematic of the experimental setup and the relevant Rydberg three-level ladder diagram are shown in Figs.~1~(a) and (b). The experiments are performed in a cylindrical room-temperature cesium vapor cell that is 50~mm long and has a 20-mm diameter. The cell is suspended between two parallel aluminum plate electrodes that are separated by $d=28$~mm. The EIT coupling-laser and probe-laser beams are overlapped and counter-propagated along the centerline of the cell (propagation direction along the $\hat{z}$-axis). The coupling and probe lasers have the same linear polarization in the $xy$ plane. The angle $\theta$ between the laser polarizations and the RF field (which points along the $\hat{y}$-axis) is varied by rotating the polarization of the laser beams with $\lambda$/2 plates, as seen in Fig.~1~(a). The weak EIT probe beam (central Rabi frequency $\Omega_p$ = 2$\pi \times $9.2~MHz and $1/e^{2}$ waist $w_0 = 75~\mu$m) has a wavelength $\lambda_{p}$ = 852~nm, and is frequency-locked to the transition $|6 S_{1/2}, F=4\rangle $ $\to$ $|6P_{3/2}, F'=5\rangle$, as shown in Fig.~1~(b). The coupling beam (central Rabi frequency $\Omega_c$ = 2$\pi \times $7.2~MHz for 60$D_{5/2}$ and $1/e^{2}$ waist $w_0= 95~\mu$m) is provided by a commercial laser (Toptica TA-SHG110), has a  wavelength of 510~nm and a linewidth of 1~MHz, and is scanned over a range of 1.5~GHz through the $|6P_{3/2}, F'=5\rangle \to |nD_J\rangle$ Rydberg transition. The EIT signal is observed by measuring the transmission of the probe laser using a photodiode (PD) after a dichroic mirror (DM). An auxiliary RF-free EIT reference setup (not shown, but similar to the one sketched in Fig. 1~(a)) is operated with the same lasers as the main setup. The auxiliary EIT signal is employed to locate the 0-detuning frequency reference point for all EIT spectra we show; it allows us to correct for small frequency drifts of the coupling laser.

The RF voltage amplitude, $V$, provided by a function generator (Tektronix AFG3102), is applied to the electrodes as shown in Fig.~1~(a), and the RF electric-field vector, $\bf{E}$, points along $\hat{y}$ (blue arrow in Fig.~1~(a)). The RF frequency is fixed, $\omega_{RF}$ = 2$\pi \times $100~MHz, and the RF field amplitude, $E$, is varied by changing $V$. The RF-field AC-shifts the Rydberg levels and generates even-order modulation sidebands (see Fig.~1~(b)). The RF field amplitude $E$ is approximately uniform within the atom-field interaction volume. Using a finite-element calculation, we have determined that the average electric field in the atom-field interaction region is $95\%$ of the field that would be present under absence of the dielectric glass cell ({\sl{i.e.}} the glass shields $5\%$ of the field). The glass cell further gives rise to an $\approx \pm 1.5\%$ field inhomogeneity along the beam paths within the cell.

The RF transmission line between the source and the cell has unavoidable standing-wave effects. While the standing-wave effect is hard to model due to the details of the experimental setup, which are fairly complex from the viewpoint of RF field modeling, the setup still constitutes a linear transmission system. Therefore, for any given frequency and fixed arrangement of the wiring and the electromagnetic boundary conditions, the magnitude of the voltage amplitude that occurs on the RF field plates follows $V_C = t V$, where $t$ is a frequency-dependent transmission factor that is specific to the details of the RF transmission line. As discussed in detail in Section~\ref{sec:measurements}, we use the atom-based field measurement method to determine the transmission factor to be $t = 1.326 \pm 0.007$. The average RF electric-field amplitude, $E$, averaged over the atom-field interaction zone inside the cell, is then related to the known voltage amplitude, $V$, generated by the source via $E = (V_C/d) \times 0.95 = (V/d) \times 0.95 \times t$ ($d$ is the distance between the field plates). In this relation, the only factor that is difficult to determine is the transmission factor $t$~\cite{scopenote}. The experiment described in this paper represents a good example of how the atom-based field measurement method allows one to measure $t$ and to thereby calibrate RF electric fields.

\section{Rydberg-atom-based characterization of an RF field}
\label{sec:measurements}

\subsection{Electric-field calibration}
\label{subsec:spectroscopy}

\begin{figure}[thb]
\centering
\includegraphics[width=0.45\textwidth]{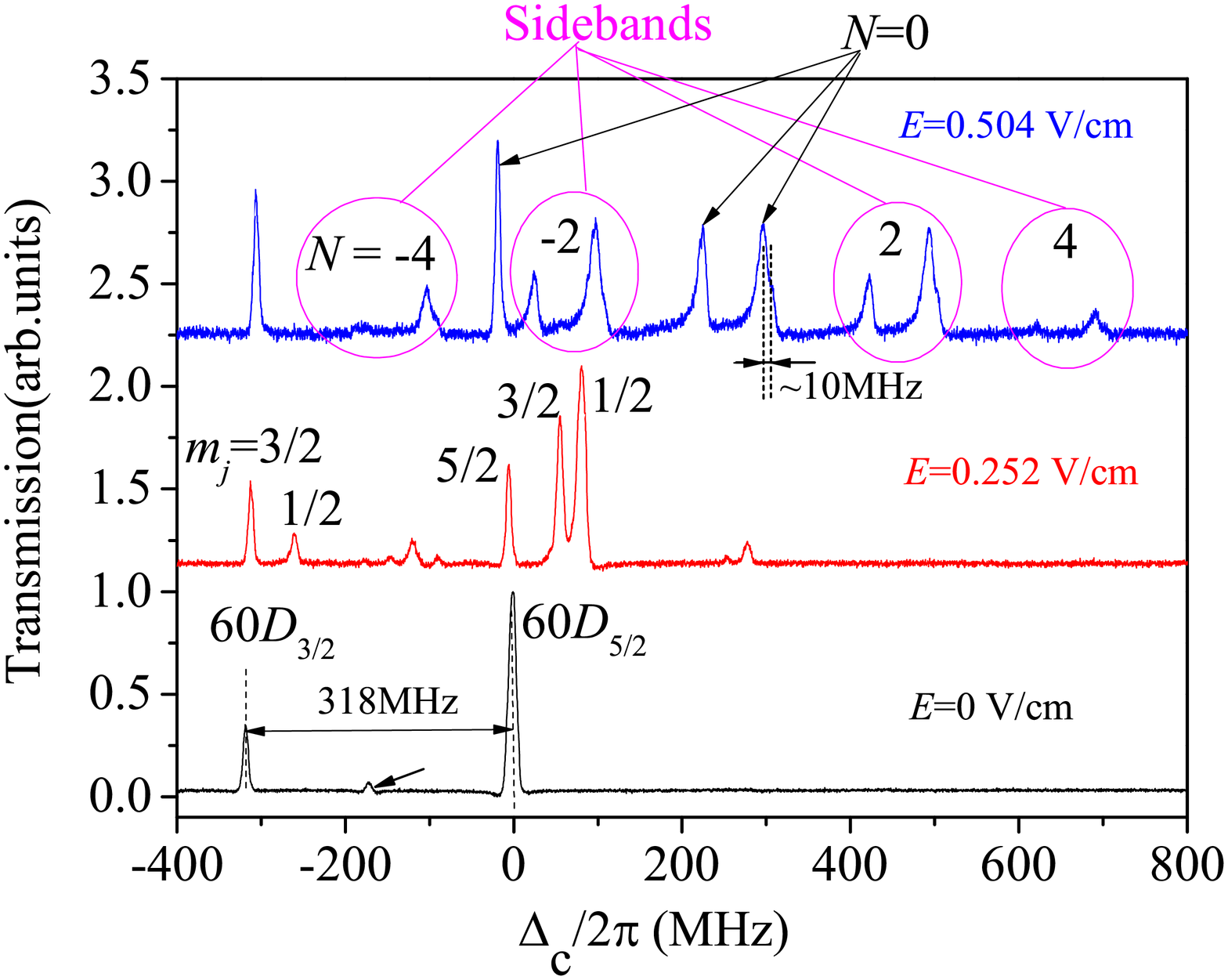}
\caption{Measured Rydberg-EIT spectra for an applied RF field with $\omega_{RF}$ = 2$\pi\times$100~MHz and the indicated amplitudes $E=0$~V/cm (bottom), 0.252~V/cm (middle) and 0.504~V/cm (top), respectively. The main peak at 0 detuning in the field-free spectrum corresponds to the $\vert 6 S_{1/2} F=4 \rangle$ $\rightarrow$ $\vert 6P_{3/2} F'=5 \rangle$ $\rightarrow$ $\vert 60D_{5/2} \rangle$ cascade EIT. The small peak at -168~MHz (small arrow) originates in the intermediate-state hyperfine level $|6P_{3/2}, F'=4\rangle$~\cite{Jiao2016}. The peak at -318~MHz is the $60D_{3/2}$ EIT line. The peaks within the magenta circles are RF-induced sidebands of the indicated orders.}
\end{figure}

In Fig.~2, we show Rydberg-EIT spectra for the $60D_J$ states for $\theta=30^{\circ}$ without RF field (bottom curve) and with the indicated RF fields (upper pair of curves).
The bottom EIT spectrum is obtained with the RF-free reference setup. The $60D_{5/2}$ main peak in the reference spectrum defines the 0-detuning position. Since the value of the $60D_J$ fine structure splitting (318-MHz arrow in Fig.~2) is well known, the spacing between the zero-field $60D_J$ fine structure components is used to calibrate the detuning axis. The top two curves show EIT spectra for applied RF field strengths $E=0.252$~V/cm and 0.504~V/cm. The $E=0.252$~V/cm plot illustrates the RF-induced AC Stark shifts in weak RF-fields. The degeneracy between the $m_j$ = 1/2, 3/2, and 5/2 magnetic substates of the $60D_J$ levels becomes lifted. (The quantization axis for $m_j$ is the direction of ${\bf E}$ in Fig.~1~(a).) Since the RF field frequency is much lower than the Kepler frequency (35~GHz for Cs 60$D$), the AC shifts in weak RF fields are near-identical with $- \alpha_{DC} E^2_{RMS}/2$, where $\alpha_{DC}$ are the DC polarizabilities of the $\vert 60D_{J}, m_j \rangle$ states, and $E_{RMS} = E/\sqrt{2}$ is the RF root-mean-square field. This has been verified with a DC Stark shift calculation (not shown).
At higher fields, RF-induced even-harmonic sidebands for $N=\pm2,\pm4$ appear, which are marked with magenta circles in the top curve of Fig.~2. The sidebands come in pairs, the lower-frequency component has $m_j=3/2$, the higher-frequency one $m_j=1/2$. The lines that do not shift much throughout Fig.~2 are the $\vert 60D_{J}, m_j=J \rangle$ states; these have near-zero polarizability.
The AC shifts and sideband separations are on the same order as the fine structure splitting of 60$D_J$. This similarity in energy scales is important because it gives rise to the level crossings in the Floquet maps discussed below.

\begin{figure*}[thb]
\centering
\includegraphics[width=1.0\textwidth]{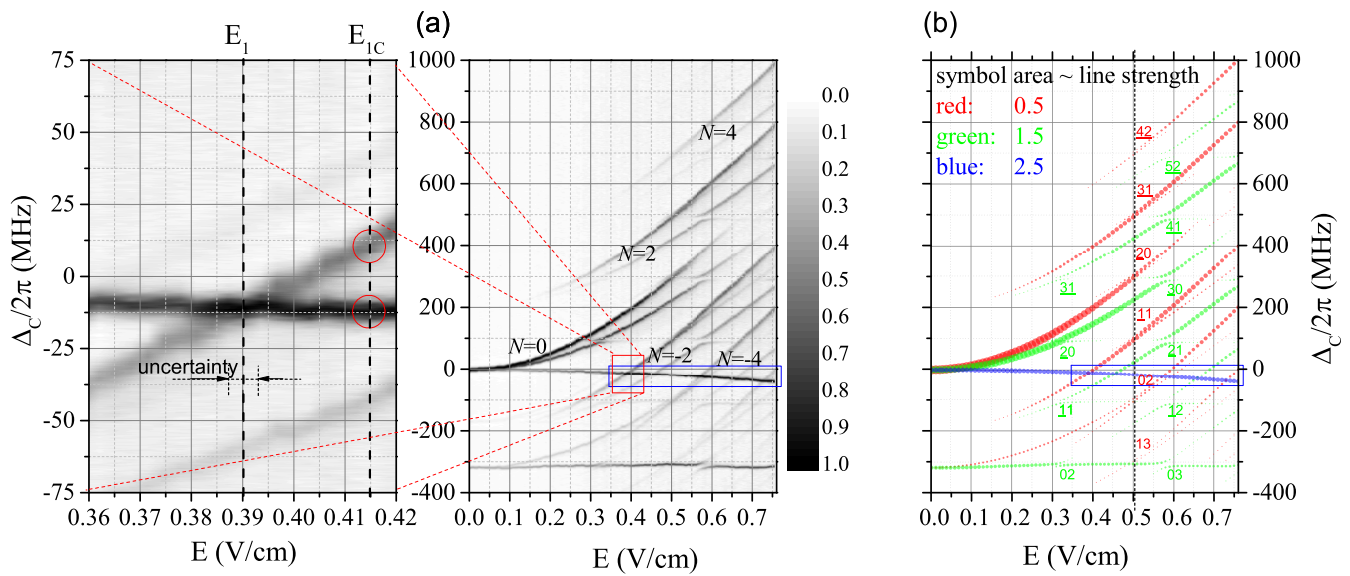}
\caption{(a) Measurement of a Cs $60D_J$ Floquet map for $\theta=30^\circ$ as a function of RF electric-field $E$ with $\omega_{RF}$ = 2$\pi\times$100~MHz. EIT line strength is displayed on a linear gray scale. As the field strength increases, the levels $\vert 60D_J, m_j \rangle$ become AC-shifted with $m_j$-dependent polarizabilites, and even level-modulation sidebands appear. The $N$-labels show sideband orders of the $\vert 60D_{5/2}, m_j=1/2, 3/2 \rangle$ states. The $\vert 60D_{5/2}, m_j=5/2 \rangle$ state has near-zero polarizability and crosses with several $m_{j}$ =3/2, 1/2 bands. These crossings are used for atom-based RF-field calibration (see text). The panel on the left shows a zoom-in of the crossing at $E_{1}$=0.390~V/cm, indicated by the red square. (b) Calculated Floquet map, with symbol areas proportional to EIT line strength. The labels of the avoided crossings are explained in the text. The rectangular boxes enclose the six level crossings we use for field calibration.}
\end{figure*}

We have performed a series of measurements such as in Fig.~2 over an $E$-field range of 0 to 0.76~V/cm, in steps of 0.006~V/cm. We have assembled the RF-EIT spectra in a Floquet map, shown Fig.~3~(a). At fields $E \lesssim 0.2~$V/cm the $m_j$-sublevels shift and split due to the $m_J$-dependent quadratic AC Stark effect. The even-harmonic level modulation sidebands, labeled $N=\pm2,\pm4$, begin to appear when the RF field is increased further (also see previous work~\cite{Bason2010,Jiao2016}). To match the measured EIT spectra with theory, we numerically calculate Rydberg EIT spectra using Floquet theory, with results shown in Fig.~3~(b). For details of the Floquet calculation see~\cite{Jiao2016,Anderson2014,Anderson2016}.

A central point of the present work is that the $\vert 60D_{5/2}, m_j=5/2 \rangle$ level, which has near-zero polarizability and AC Stark shift, undergoes a series of crossings with the $m_{j}$ =1/2, 3/2 modulation sidebands. The crossings are exact because the linearly polarized RF field does not mix quantum states of different $m_j$. The crossings can be measured with about $1\%$ precision. As an example, in Fig.~3~(a) we show a zoom-in of the first level crossing. The crossing is centered at $E_1=0.390$~V/cm, with an estimated uncertainty of $\pm 0.004$~V/cm, corresponding to a relative uncertainty of $\pm 1\%$. The uncertainty is mostly attributed to the intrinsic EIT linewidth, which increases with increasing coupling and probe Rabi frequencies. Laser linewidths and interaction-time broadening also contribute to the observed linewidths.

\begin{table}
    \caption{The columns show, in that order, the crossing number, the calculated electric field $E$ for the crossing, the experimental electric field $E_0 = 0.95 \times V/d$ the atoms would be exposed to for an RF amplitude transmission factor $t=1$, and the transmission factor $t = E/E_0$.}
    \begin{center}
\begin{tabular}{|c|c|c|c|}
  \hline
  crossing\# $i$ & $E_i$~(V/cm) & $E_{0,i}$~(V/cm) & $t_i$  \\
  \hline
  1 & 0.390 & 0.2964 & 1.3158 \\
  2 & 0.457 & 0.3449 & 1.3252 \\
  3 & 0.576 & 0.4323 & 1.3326 \\
  4 & 0.594 & 0.4456 & 1.3332 \\
  5 & 0.657 & 0.4978 & 1.3198 \\
  6 & 0.732 & 0.5510 & 1.3285 \\
  \hline
\end{tabular}
    \end{center}
\end{table}
In Fig.~3 six such crossings are visible within the rectangular boxes. With the RF-source voltage amplitudes $V_i$ at which the crossings are observed, and recalling that the glass cell shields $5\%$ of the electric field from the atoms, the electric field the atoms would experience for an amplitude transmission factor of 1  would be $E_{0,i} = 0.95 \times V_i/d$. The ratios between the known (theoretical) electric fields where the crossings actually occur, $E_i$, and the $E_{0,i}$ yield six readings for the amplitude transmission factor, $t_i = E_i / E_{0,i}$. In Table~1 it is seen that the $t_i$ have a very small spread and do not exhibit a systematic trend from low to high field. The average, $t=1.326 \pm 0.007$, is the desired calibration factor for the experimental electric-field axis. The $x-$axis in Fig.~3~(a) shows the calibrated experimental electric field, $E = t E_0 = t \times 0.95 \times V/d$, with voltage amplitude $V$ at the source. The overall relative uncertainty of the atom-based RF-field calibration performed in this experiment is $\pm 0.5\%$, similar to what has been obtained in~\cite{Miller2016} and about an order of magnitude better than in traditional RF field calibration~\cite{Hill1990,Matloubi1993}. The use of narrow-band coupling and probe lasers, lower Rabi frequencies, and larger-diameter laser beams is expected to reduce the uncertainty to considerably smaller values.

We note that the calibration uncertainty achieved in this work is based on matching experimental and calculated spectroscopic data at the locations of a series of six isolated level crossings that all occur within a narrow spectral range of less than 50~MHz width (see rectangular boxes in Fig.~3). Hence, a fairly small amount of spectroscopic data suffices for the presented atom-based RF field calibration. From Fig.~3 it is obvious that this advantage traces back to a specific feature of cesium $nD$ states, namely that these states offer a mix of magnetic sublevels with near-zero and large AC polarizabilities.  RF-dressed Rydberg-EIT spectra of rubidium atoms do not present a similar advantage~\cite{Miller2016}.

In Fig.~3 we further observe three series of avoided crossings, which are due to an RF-sideband of the $J=3/2$ level intersecting with an RF-sideband of the $J=5/2$ level. The first number in the avoided-crossing labels in Fig.~3~(b) shows the number of RF photon pairs taken from the RF field to access the $J=3/2$ band, while the second shows the number of RF photon pairs taken from the RF field to access the $J=5/2$ band. Negative RF photon numbers, indicated by underbars, correspond to stimulated RF-photon emission. The coupling between the intersecting $J=3/2$ and $J=5/2$ bands is a two-RF-photon Raman process in which the atom absorbs and re-emits an RF photon while changing $J$ from $3/2$ to $5/2$, or vice versa. This is a second-order electric-dipole transition, which, for the given polarization, has selection rules $\Delta L = 0, 2$ and $\Delta m_j = 0$. In Fig.~3, three series of avoided crossings that satisfy these selection rules are visible, one for $m_j = 1/2$ and two for $m_j = 3/2$.
Each series has a fixed $E$-value and consists of copies of the same avoided crossing along the $\Delta_c$-axis, in steps of 200~MHz.
The $m_j = 3/2$ series are particularly easy to spot because one of the two intersecting Floquet states has near-zero polarizability. The Raman coupling causing the avoided crossings equals the minimal avoided-crossing gap size. For fixed Floquet-state wavefunction, the Raman coupling strength should scale as $E^2$. For the $m_j = 3/2$ avoided crossings at 0.319~V/cm we observe a coupling strength of 8.6~MHz, while those at 0.579~V/cm have a coupling strength of 19.3~MHz. The coupling-strength ratio, which is 2.2, is somewhat smaller than the $E^2$-ratio, which is 3.3. The deviation indicates a moderate variation of the Floquet-state wavefunctions between 0.319~V/cm and 0.579~V/cm (which is expected).
From a field-calibration point of view, the avoided crossings and other details in the spectra could be used to further reduce the uncertainty in the atom-based RF-field calibration factor $t$, which is planned in future work.
Comparing the cesium and rubidium level structures, it is again noteworthy that cesium offers a combination of $m_j$-dependent polarizabilities that is particularly favorable for this purpose.

In the top curve in Fig.~2 it is noted that the $\vert 60D_{5/2}, m_j = 5/2 \rangle$ and $\vert 60D_{3/2}, m_j = 3/2 \rangle$ Floquet states are narrow and symmetric, whereas the other Floquet lines are much wider and are asymmetrically broadened. Further, the $m_j=1/2$-lines exhibit a shoulder on the high-frequency side (see $\approx 10~$MHz marker in Fig.~2), while the $m_j=3/2$-lines have no shoulder. The scan in the top curve of Fig~2 also corresponds to the vertical dashed line in Fig.~3~(b). Close inspection of Fig.~3~(b) reveals that the shoulders of the $m_j=1/2$-lines are due to the series of narrow avoided crossings between Floquet states in the $m_j=1/2$-manifold. The shoulders correspond to the weaker, higher-frequency component of the crossing. The asymmetric line broadening of the wide lines is due to the $3 \%$ full-width variation of the RF field within the atom-field interaction zone. For instance, for the $m_j=3/2$-lines we estimate for the inhomogeneous linewidth $\Delta W = 3 \% E \frac{dW}{dE} \approx 13~$MHz, which is close to the observed width of $\sim 15$~MHz. (The $m_j=1/2$-lines are also inhomogeneously broadened, but we do not give a broadening estimate for those lines because of the interference with the mentioned avoided crossing.)

\subsection{RF polarization measurement}
\label{subsec:polarization}

\begin{figure} [thb]
\centering
\includegraphics[width=0.4\textwidth]{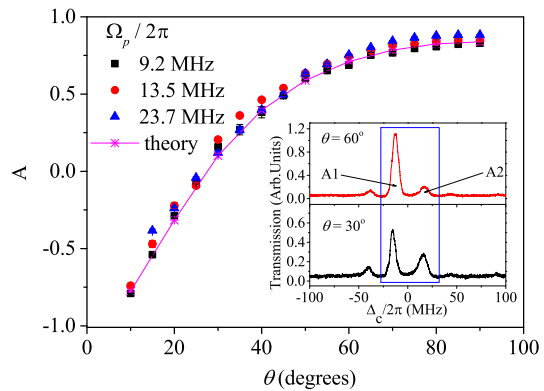}
\caption{Experimental data (symbols) and calculations (line) for the line strength ratio $A=(A1-A2)/(A1+A2)$ defined in the text at an RF field of $E_{1C}$=0.415~V/cm, as a function of polarization angle $\theta$. The data are for the indicated values of the probe Rabi frequency. The inset shows sample EIT spectra for $\theta = 30^\circ$ and $60^\circ$.}
\end{figure}

Rydberg-EIT spectra generally depend on the laser polarizations~\cite{McGloin2000,Bao2016}.
This also applies to RF-modulated Rydberg EIT spectra. Here, we study the dependence of line-strength ratios on the angle $\theta$ between the RF-field and the polarization of the laser beams (both laser beams are linearly polarized, and the polarizations are parallel to each other; see Fig.~1). For a line-strength comparison of Floquet levels of different $m_j$ it is advantageous to choose an electric field close to one of the exact crossings discussed above, because the two lines of interest will then appear in close proximity to each other, allowing for a rapid measurement.
Additionally, since states with different $m_j$ do not mix, the line-strength measurements are robust against small variations of the RF electric field.
As an example, the inset in Fig.~4 we show the RF EIT spectra for $\theta=30^\circ$  (lower curve) and $60^\circ$ (upper curve) at an RF field $E_{1C}$= 0.415~V/cm, marked with a dashed line in the left panel in Fig.~3~(a). The two peaks labeled A1 and A2 within the blue square in the inset in Fig.~4 correspond to the Floquet levels  marked with red circles in Fig.~3~(a). The peak A1, which corresponds to the $N=0$ RF band of the $\vert J=5/2, m_j=5/2 \rangle$ Floquet state, increases with the angle $\theta$, whereas the peak A2, which corresponds to the $N=2$ RF band of the $\vert J=5/2, m_j=1/2 \rangle$ state, decreases with $\theta$. To quantify this polarization-angle dependence, we introduce the parameter $A(\theta) = (A1-A2)/(A1+A2)$, where $A1 (A2)$ represent the respective areas of Gaussian peaks obtained from double-Gaussian fits to the spectra at angle $\theta$.
Since the intersecting lines have different differential dipole moments, it is important to use the areas and not the peak heights (see discussion in the last paragraph in Sec.~\ref{subsec:spectroscopy}). Figure~4 shows $A$ as a function of the $\theta$ at $E_{1C}$=0.415~V/cm for the indicated probe laser Rabi frequencies, together with the corresponding line strength ratio obtained from Floquet calculations (the Floquet calculation yields line strengths valid for the case of low saturation, $\Omega_p < \Gamma_2$). We find excellent agreement between the measurements and calculations for $\Omega_p$ = $2\pi \times$ 9.2~MHz. Curves such as in Fig.~4 can be used to measure the polarization of an RF-field with unknown linear polarization. The angle uncertainty can be estimated as $\Delta \theta = \vert \Delta A \vert / \frac{dA}{d \theta}$, where  $\Delta A$ is the difference between experimental and calculated values of $A(\theta)$, and $\frac{dA}{d \theta}$ is the derivative of the calculated curve. For the lowest-power case in Fig.~4, straightforward analysis shows angle uncertainties below $2^\circ$ for $\theta < 60^\circ$. In the domain $60^\circ <  \theta < 90^\circ$ the angle uncertainty gradually increases from $2^\circ$ to $7^\circ$ because the derivative becomes small. We note that this method of polarization measurement has the advantage of being both simple and very fast, since the areas of only two lines need to be measured.
At the expense of reduced acquisition speed, the uncertainty could be improved by measuring line-strength ratios of multiple line pairs and by averaging over a number of spectra.

The data for higher probe Rabi frequencies in Fig.~4 show a more significant deviation from the calculated curve. This is not unexpected, because the calculation is for negligible saturation of the probe transition, whereas the data in Fig.~4 vary between moderate and strong saturation of the probe transition. In addition to saturation broadening effects, there may also be optical-pumping effects~\cite{Zhang2017} that could affect the line strength ratio. This is beyond the scope of the present work.

\section{Conclusion}

We have demonstrated a rapid and robust atom-based method to calibrate the electric field and to measure the polarization of
a 100~MHz RF field, using Rydberg EIT in a room-temperature cesium vapor cell as an all-optical field probe. The EIT spectra exhibit RF-field-induced AC Stark shifts, splittings and even-order level modulation sidebands. A series of exact Floquet level intersections that are specific to cesium Rydberg atoms have been used for calibrating the RF electric field with an uncertainty of $\pm 0.5\%$. The dependence of the Rydberg-EIT spectra on the polarization angle of the RF field has been studied. Our analysis of certain line-strength ratios has led into a convenient method to determine the polarization of the RF electric field. The Rydberg-EIT spectroscopy presented here could be applied to atom-based, antenna-free calibration of RF electric fields and polarization measurement. It is anticipated that an extended analysis of all exact and avoided crossings as well as other spectroscopic features will significantly lower the calibration uncertainty.
Future work involving narrow-band laser sources, miniature spectroscopic cells as well as improved spectroscopic methods (lower Rabi frequencies, wider probe and coupler beams) are expected to further reduce the calibration uncertainty.

The work was supported by NNSF of China (Grants Nos. 11274209, 61475090, 61475123), Changjiang Scholars and Innovative Research Team in University of Ministry of Education of China (Grant No. IRT13076), the State Key Program of National Natural Science of China (Grant No. 11434007), and Research Project Supported by Shanxi Scholarship Council of China (2014-009). GR acknowledges support by the NSF (PHY-1506093) and BAIREN plan of Shanxi province.


\begin{thebibliography}{27}%
\makeatletter
\providecommand \@ifxundefined [1]{%
 \@ifx{#1\undefined}
}%
\providecommand \@ifnum [1]{%
 \ifnum #1\expandafter \@firstoftwo
 \else \expandafter \@secondoftwo
 \fi
}%
\providecommand \@ifx [1]{%
 \ifx #1\expandafter \@firstoftwo
 \else \expandafter \@secondoftwo
 \fi
}%
\providecommand \natexlab [1]{#1}%
\providecommand \enquote  [1]{``#1''}%
\providecommand \bibnamefont  [1]{#1}%
\providecommand \bibfnamefont [1]{#1}%
\providecommand \citenamefont [1]{#1}%
\providecommand \href@noop [0]{\@secondoftwo}%
\providecommand \href [0]{\begingroup \@sanitize@url \@href}%
\providecommand \@href[1]{\@@startlink{#1}\@@href}%
\providecommand \@@href[1]{\endgroup#1\@@endlink}%
\providecommand \@sanitize@url [0]{\catcode `\\12\catcode `\$12\catcode
  `\&12\catcode `\#12\catcode `\^12\catcode `\_12\catcode `\%12\relax}%
\providecommand \@@startlink[1]{}%
\providecommand \@@endlink[0]{}%
\providecommand \url  [0]{\begingroup\@sanitize@url \@url }%
\providecommand \@url [1]{\endgroup\@href {#1}{\urlprefix }}%
\providecommand \urlprefix  [0]{URL }%
\providecommand \Eprint [0]{\href }%
\providecommand \doibase [0]{http://dx.doi.org/}%
\providecommand \selectlanguage [0]{\@gobble}%
\providecommand \bibinfo  [0]{\@secondoftwo}%
\providecommand \bibfield  [0]{\@secondoftwo}%
\providecommand \translation [1]{[#1]}%
\providecommand \BibitemOpen [0]{}%
\providecommand \bibitemStop [0]{}%
\providecommand \bibitemNoStop [0]{.\EOS\space}%
\providecommand \EOS [0]{\spacefactor3000\relax}%
\providecommand \BibitemShut  [1]{\csname bibitem#1\endcsname}%
\let\auto@bib@innerbib\@empty
\bibitem [{\citenamefont {Heavner}\ \emph {et~al.}(2014)\citenamefont
  {Heavner}, \citenamefont {Donley}, \citenamefont {Levi}, \citenamefont
  {Costanzo}, \citenamefont {Parker}, \citenamefont {Shirley}, \citenamefont
  {Ashby}, \citenamefont {Barlow},\ and\ \citenamefont
  {Jefferts}}]{Heavner2014}%
  \BibitemOpen
  \bibfield  {author} {\bibinfo {author} {\bibfnamefont {T. P.}\ \bibnamefont
  {Heavner}}, \bibinfo {author} {\bibfnamefont {E.A.}\ \bibnamefont {Donley}},
  \bibinfo {author} {\bibfnamefont {F.}~\bibnamefont {Levi}}, \bibinfo {author}
  {\bibfnamefont {G.}~\bibnamefont {Costanzo}}, \bibinfo {author}
  {\bibfnamefont {T. E.}\ \bibnamefont {Parker}}, \bibinfo {author}
  {\bibfnamefont {J. H.}\ \bibnamefont {Shirley}}, \bibinfo {author}
  {\bibfnamefont {N.}~\bibnamefont {Ashby}}, \bibinfo {author} {\bibfnamefont
  {S.}~\bibnamefont {Barlow}}, \ and\ \bibinfo {author} {\bibfnamefont {S. R. }\
  \bibnamefont {Jefferts}},\ }\bibfield  {title} {\enquote {\bibinfo {title}
  {First accuracy evaluation of NIST-F2},}\ }\href
  {http://dx.doi.org/10.1088/0026-1394/51/3/174} {\bibfield  {journal}
  {\bibinfo  {journal} {Metrologia}\ }\textbf {\bibinfo {volume} {51}},\
  \bibinfo {pages} {174 -- 82} (\bibinfo {year} {2014})}\BibitemShut {NoStop}%
\bibitem [{\citenamefont {Ludlow}\ \emph {et~al.}(2015)\citenamefont {Ludlow},
  \citenamefont {Boyd}, \citenamefont {Ye}, \citenamefont {Peik},\ and\
  \citenamefont {Schmidt}}]{Ludlow2015}%
  \BibitemOpen
  \bibfield  {author} {\bibinfo {author} {\bibfnamefont {A. D. }\ \bibnamefont
  {Ludlow}}, \bibinfo {author} {\bibfnamefont {M. M. }\ \bibnamefont {Boyd}},
  \bibinfo {author} {\bibfnamefont {Jun}\ \bibnamefont {Ye}}, \bibinfo {author}
  {\bibfnamefont {E.}~\bibnamefont {Peik}}, \ and\ \bibinfo {author}
  {\bibfnamefont {P. O. }\ \bibnamefont {Schmidt}},\ }\bibfield  {title}
  {\enquote {\bibinfo {title} {Optical atomic clocks},}\ }\href
  {http://dx.doi.org/10.1103/RevModPhys.87.637} {\bibfield  {journal} {\bibinfo
   {journal} {Rev. Mod. Phys.}\ }\textbf {\bibinfo {volume} {87}},\ \bibinfo
  {pages} {637 -- 701} (\bibinfo {year} {2015})}\BibitemShut {NoStop}%
\bibitem [{\citenamefont {Savukov}\ \emph {et~al.}(2005)\citenamefont
  {Savukov}, \citenamefont {Seltzer}, \citenamefont {Romalis},\ and\
  \citenamefont {Sauer}}]{Savukov2005}%
  \BibitemOpen
  \bibfield  {author} {\bibinfo {author} {\bibfnamefont {I. M. }\ \bibnamefont
  {Savukov}}, \bibinfo {author} {\bibfnamefont {S.J.}\ \bibnamefont {Seltzer}},
  \bibinfo {author} {\bibfnamefont {M. V. }\ \bibnamefont {Romalis}}, \ and\
  \bibinfo {author} {\bibfnamefont {K. L. }\ \bibnamefont {Sauer}},\ }\bibfield
  {title} {\enquote {\bibinfo {title} {Tunable atomic magnetometer for
  detection of radio-frequency magnetic fields},}\ }\href
  {http://dx.doi.org/10.1103/PhysRevLett.95.063004} {\bibfield  {journal}
  {\bibinfo  {journal} {Phys. Rev. Lett.}\ }\textbf {\bibinfo {volume} {95}},\
  \bibinfo {pages} {063004--1--4} (\bibinfo {year} {2005})}\BibitemShut
  {NoStop}%
\bibitem [{\citenamefont {Patton}\ \emph {et~al.}(2012)\citenamefont {Patton},
  \citenamefont {Versolato}, \citenamefont {Hovde}, \citenamefont {Corsini},
  \citenamefont {Higbie},\ and\ \citenamefont {Budker}}]{Patton2012}%
  \BibitemOpen
  \bibfield  {author} {\bibinfo {author} {\bibfnamefont {B.}~\bibnamefont
  {Patton}}, \bibinfo {author} {\bibfnamefont {O. O.}\ \bibnamefont
  {Versolato}}, \bibinfo {author} {\bibfnamefont {D. C.}\ \bibnamefont {Hovde}},
  \bibinfo {author} {\bibfnamefont {E.}~\bibnamefont {Corsini}}, \bibinfo
  {author} {\bibfnamefont {J. M.}\ \bibnamefont {Higbie}}, \ and\ \bibinfo
  {author} {\bibfnamefont {D.}~\bibnamefont {Budker}},\ }\bibfield  {title}
  {\enquote {\bibinfo {title} {A remotely interrogated all-optical 87Rb
  magnetometer},}\ }\href {http://dx.doi.org/10.1063/1.4747206} {\bibfield
  {journal} {\bibinfo  {journal} {Appl. Phys. Lett.}\ }\textbf {\bibinfo
  {volume} {101}},\ \bibinfo {pages} {083502--1--4} (\bibinfo {year}
  {2012})}\BibitemShut {NoStop}%
\bibitem [{\citenamefont {Mohapatra}\ \emph {et~al.}(2007)\citenamefont
  {Mohapatra}, \citenamefont {Jackson},\ and\ \citenamefont
  {Adams}}]{Mohapatra2007}%
  \BibitemOpen
  \bibfield  {author} {\bibinfo {author} {\bibfnamefont {A.~K.}\ \bibnamefont
  {Mohapatra}}, \bibinfo {author} {\bibfnamefont {T.~R.}\ \bibnamefont
  {Jackson}}, \ and\ \bibinfo {author} {\bibfnamefont {C.~S.}\ \bibnamefont
  {Adams}},\ }\bibfield  {title} {\enquote {\bibinfo {title} {Coherent optical
  detection of highly excited Rydberg states using electromagnetically induced
  transparency},}\ }\href {\doibase 10.1103/PhysRevLett.98.113003} {\bibfield
  {journal} {\bibinfo  {journal} {Phys. Rev. Lett.}\ }\textbf {\bibinfo
  {volume} {98}},\ \bibinfo {pages} {113003--1--4} (\bibinfo {year}
  {2007})}\BibitemShut {NoStop}%
\bibitem [{\citenamefont {T.F.Gallagher}(1994)}]{Gallagher1994}%
  \BibitemOpen
  \bibfield  {author} {\bibinfo {author} {\bibnamefont {T. F. Gallagher}},\
  }\href@noop {} {\emph {\bibinfo {title} {Rydberg Atoms}}}\ (\bibinfo
  {publisher} {Cambridge University Press},\ \bibinfo {address} {New York, NY,
  USA},\ \bibinfo {year} {1994})\BibitemShut {NoStop}%
\bibitem [{\citenamefont {Sedlacek}\ \emph {et~al.}(2012)\citenamefont
  {Sedlacek}, \citenamefont {Schwettmann}, \citenamefont {K\"ubler},
  \citenamefont {L\"ow}, \citenamefont {Pfau},\ and\ \citenamefont
  {Shaffer}}]{Sedlacek2012}%
  \BibitemOpen
  \bibfield  {author} {\bibinfo {author} {\bibfnamefont {J.~A.}\ \bibnamefont
  {Sedlacek}}, \bibinfo {author} {\bibfnamefont {A.}~\bibnamefont
  {Schwettmann}}, \bibinfo {author} {\bibfnamefont {H.}~\bibnamefont
  {K\"ubler}}, \bibinfo {author} {\bibfnamefont {R.}~\bibnamefont {L\"ow}},
  \bibinfo {author} {\bibfnamefont {T.}~\bibnamefont {Pfau}}, \ and\ \bibinfo
  {author} {\bibfnamefont {James~P.}\ \bibnamefont {Shaffer}},\ }\bibfield
  {title} {\enquote {\bibinfo {title} {Microwave electrometry with Rydberg
  atoms in a vapour cell using bright atomic resonances},}\ }\href {\doibase
  10.1038/nphys2423} {\bibfield  {journal} {\bibinfo  {journal} {Nat. Phys.}\
  }\textbf {\bibinfo {volume} {8}},\ \bibinfo {pages} {819--824} (\bibinfo
  {year} {2012})}\BibitemShut {NoStop}%
\bibitem [{\citenamefont {Sedlacek}\ \emph {et~al.}(2013)\citenamefont
  {Sedlacek}, \citenamefont {Schwettmann}, \citenamefont {K\"ubler},\ and\
  \citenamefont {Shaffer}}]{Sedlacek2013}%
  \BibitemOpen
  \bibfield  {author} {\bibinfo {author} {\bibfnamefont {J.~A.}\ \bibnamefont
  {Sedlacek}}, \bibinfo {author} {\bibfnamefont {A.}~\bibnamefont
  {Schwettmann}}, \bibinfo {author} {\bibfnamefont {H.}~\bibnamefont
  {K\"ubler}}, \ and\ \bibinfo {author} {\bibfnamefont {J.~P.}\ \bibnamefont
  {Shaffer}},\ }\bibfield  {title} {\enquote {\bibinfo {title} {Atom-based
  vector microwave electrometry using rubidium Rydberg atoms in a vapor
  cell},}\ }\href {\doibase 10.1103/PhysRevLett.111.063001} {\bibfield
  {journal} {\bibinfo  {journal} {Phys. Rev. Lett.}\ }\textbf {\bibinfo
  {volume} {111}},\ \bibinfo {pages} {063001--1--5} (\bibinfo {year}
  {2013})}\BibitemShut {NoStop}%
\bibitem [{\citenamefont {Fan}\ \emph {et~al.}(2015)\citenamefont {Fan},
  \citenamefont {Kumar}, \citenamefont {Sedlacek}, \citenamefont {K\"ubler},
  \citenamefont {Karimkashi},\ and\ \citenamefont {Shaffer}}]{Fan2015}%
  \BibitemOpen
  \bibfield  {author} {\bibinfo {author} {\bibfnamefont {H.}~\bibnamefont
  {Fan}}, \bibinfo {author} {\bibfnamefont {S.}~\bibnamefont {Kumar}}, \bibinfo
  {author} {\bibfnamefont {J.}~\bibnamefont {Sedlacek}}, \bibinfo {author}
  {\bibfnamefont {H.}~\bibnamefont {K\"ubler}}, \bibinfo {author}
  {\bibfnamefont {S.}~\bibnamefont {Karimkashi}}, \ and\ \bibinfo {author}
  {\bibfnamefont {J.P.}\ \bibnamefont {Shaffer}},\ }\bibfield  {title}
  {\enquote {\bibinfo {title} {Atom based RF electric field sensing},}\ }\href
  {http://dx.doi.org/10.1088/0953-4075/48/20/202001} {\bibfield  {journal}
  {\bibinfo  {journal} {J. Phys. B}\ }\textbf {\bibinfo {volume} {48}},\
  \bibinfo {pages} {202001--1--16} (\bibinfo {year} {2015})}\BibitemShut
  {NoStop}%
\bibitem [{\citenamefont {Gordon}\ \emph {et~al.}(2014)\citenamefont {Gordon},
  \citenamefont {Holloway}, \citenamefont {Schwarzkopf}, \citenamefont
  {Anderson}, \citenamefont {Miller}, \citenamefont {Thaicharoen},\ and\
  \citenamefont {Raithel}}]{Gordon2014}%
  \BibitemOpen
  \bibfield  {author} {\bibinfo {author} {\bibfnamefont {J. A.}\ \bibnamefont
  {Gordon}}, \bibinfo {author} {\bibfnamefont {C. L.}\ \bibnamefont {Holloway}},
  \bibinfo {author} {\bibfnamefont {A.}~\bibnamefont {Schwarzkopf}}, \bibinfo
  {author} {\bibfnamefont {D. A.}\ \bibnamefont {Anderson}}, \bibinfo {author}
  {\bibfnamefont {S.}~\bibnamefont {Miller}}, \bibinfo {author} {\bibfnamefont
  {N.}~\bibnamefont {Thaicharoen}}, \ and\ \bibinfo {author} {\bibfnamefont
  {G.}~\bibnamefont {Raithel}},\ }\bibfield  {title} {\enquote {\bibinfo
  {title} {Millimeter wave detection via Autler-Townes splitting in rubidium
  Rydberg atoms},}\ }\href {http://dx.doi.org/10.1063/1.4890094} {\bibfield
  {journal} {\bibinfo  {journal} {Appl. Phys. Lett.}\ }\textbf {\bibinfo
  {volume} {105}},\ \bibinfo {pages} {024104--1--5} (\bibinfo {year}
  {2014})}\BibitemShut {NoStop}%
\bibitem [{\citenamefont {Barredo}\ \emph {et~al.}(2013)\citenamefont
  {Barredo}, \citenamefont {K\"ubler}, \citenamefont {Daschner}, \citenamefont
  {L\"ow},\ and\ \citenamefont {Pfau}}]{Barredo2013}%
  \BibitemOpen
  \bibfield  {author} {\bibinfo {author} {\bibfnamefont {D.}~\bibnamefont
  {Barredo}}, \bibinfo {author} {\bibfnamefont {H.}~\bibnamefont {K\"ubler}},
  \bibinfo {author} {\bibfnamefont {R.}~\bibnamefont {Daschner}}, \bibinfo
  {author} {\bibfnamefont {R.}~\bibnamefont {L\"ow}}, \ and\ \bibinfo {author}
  {\bibfnamefont {T.}~\bibnamefont {Pfau}},\ }\bibfield  {title} {\enquote
  {\bibinfo {title} {Electrical readout for coherent phenomena involving
  Rydberg atoms in thermal vapor cells},}\ }\href
  {http://dx.doi.org/10.1103/PhysRevLett.110.123002} {\bibfield  {journal}
  {\bibinfo  {journal} {Phys. Rev. Lett.}\ }\textbf {\bibinfo {volume} {110}},\
  \bibinfo {pages} {123002--1--5} (\bibinfo {year} {2013})}\BibitemShut
  {NoStop}%
\bibitem [{\citenamefont {Fan}\ \emph {et~al.}(2014)\citenamefont {Fan},
  \citenamefont {Kumar}, \citenamefont {Daschner}, \citenamefont {K\"{u}bler},\
  and\ \citenamefont {Shaffer}}]{Fan2014}%
  \BibitemOpen
  \bibfield  {author} {\bibinfo {author} {\bibfnamefont {H.~Q.}\ \bibnamefont
  {Fan}}, \bibinfo {author} {\bibfnamefont {S.}~\bibnamefont {Kumar}}, \bibinfo
  {author} {\bibfnamefont {R.}~\bibnamefont {Daschner}}, \bibinfo {author}
  {\bibfnamefont {H.}~\bibnamefont {K\"{u}bler}}, \ and\ \bibinfo {author}
  {\bibfnamefont {J.~P.}\ \bibnamefont {Shaffer}},\ }\bibfield  {title}
  {\enquote {\bibinfo {title} {Subwavelength microwave electric-field imaging
  using Rydberg atoms inside atomic vapor cells},}\ }\href {\doibase
  10.1364/OL.39.003030} {\bibfield  {journal} {\bibinfo  {journal} {Opt.
  Lett.}\ }\textbf {\bibinfo {volume} {39}},\ \bibinfo {pages} {3030--3033}
  (\bibinfo {year} {2014})}\BibitemShut {NoStop}%
\bibitem [{\citenamefont {Holloway}\ \emph
  {et~al.}(2014{\natexlab{a}})\citenamefont {Holloway}, \citenamefont {Gordon},
  \citenamefont {Schwarzkopf}, \citenamefont {Anderson}, \citenamefont
  {Miller}, \citenamefont {Thaicharoen},\ and\ \citenamefont
  {Raithel}}]{Holloway2014a}%
  \BibitemOpen
  \bibfield  {author} {\bibinfo {author} {\bibfnamefont {C. L.}\ \bibnamefont
  {Holloway}}, \bibinfo {author} {\bibfnamefont {J. A.}\ \bibnamefont {Gordon}},
  \bibinfo {author} {\bibfnamefont {A.}~\bibnamefont {Schwarzkopf}}, \bibinfo
  {author} {\bibfnamefont {D. A.}\ \bibnamefont {Anderson}}, \bibinfo {author}
  {\bibfnamefont {S. A.}\ \bibnamefont {Miller}}, \bibinfo {author}
  {\bibfnamefont {N.}~\bibnamefont {Thaicharoen}}, \ and\ \bibinfo {author}
  {\bibfnamefont {G.}~\bibnamefont {Raithel}},\ }\bibfield  {title} {\enquote
  {\bibinfo {title} {Sub-wavelength imaging and field mapping via
  electromagnetically induced transparency and Autler-Townes splitting in
  Rydberg atoms},}\ }\href {\doibase http://dx.doi.org/10.1063/1.4883635}
  {\bibfield  {journal} {\bibinfo  {journal} {Appl. Phys. Lett.}\ }\textbf
  {\bibinfo {volume} {104}},\ \bibinfo {pages} {244102--1--5} (\bibinfo {year}
  {2014}{\natexlab{a}})}\BibitemShut {NoStop}%
\bibitem [{\citenamefont {Bason}\ \emph {et~al.}(2010)\citenamefont {Bason},
  \citenamefont {Tanasittikosol}, \citenamefont {Sargsyan}, \citenamefont
  {Mohapatra}, \citenamefont {Sarkisyan}, \citenamefont {Potvliege},\ and\
  \citenamefont {Adams}}]{Bason2010}%
  \BibitemOpen
  \bibfield  {author} {\bibinfo {author} {\bibfnamefont {M.~G.}\ \bibnamefont
  {Bason}}, \bibinfo {author} {\bibfnamefont {M.}~\bibnamefont
  {Tanasittikosol}}, \bibinfo {author} {\bibfnamefont {A.}~\bibnamefont
  {Sargsyan}}, \bibinfo {author} {\bibfnamefont {A.~K.}\ \bibnamefont
  {Mohapatra}}, \bibinfo {author} {\bibfnamefont {D.}~\bibnamefont
  {Sarkisyan}}, \bibinfo {author} {\bibfnamefont {R.~M.}\ \bibnamefont
  {Potvliege}}, \ and\ \bibinfo {author} {\bibfnamefont {C.~S.}\ \bibnamefont
  {Adams}},\ }\bibfield  {title} {\enquote {\bibinfo {title} {Enhanced electric
  field sensitivity of RF-dressed Rydberg dark states},}\ }\href
  {http://dx.doi.org/10.1088/1367-2630/12/6/065015} {\bibfield  {journal}
  {\bibinfo  {journal} {New J. Phys.}\ }\textbf {\bibinfo {volume} {12}},\
  \bibinfo {pages} {065015--1--11} (\bibinfo {year} {2010})}\BibitemShut
  {NoStop}%
\bibitem [{\citenamefont {Jiao}\ \emph {et~al.}(2016)\citenamefont {Jiao},
  \citenamefont {Han}, \citenamefont {Yang}, \citenamefont {Li}, \citenamefont
  {Raithel}, \citenamefont {Zhao},\ and\ \citenamefont {Jia}}]{Jiao2016}%
  \BibitemOpen
  \bibfield  {author} {\bibinfo {author} {\bibfnamefont {Y.}~\bibnamefont
  {Jiao}}, \bibinfo {author} {\bibfnamefont {X.}~\bibnamefont {Han}}, \bibinfo
  {author} {\bibfnamefont {Z.}~\bibnamefont {Yang}}, \bibinfo {author}
  {\bibfnamefont {J.}~\bibnamefont {Li}}, \bibinfo {author} {\bibfnamefont
  {G.}~\bibnamefont {Raithel}}, \bibinfo {author} {\bibfnamefont
  {J.}~\bibnamefont {Zhao}}, \ and\ \bibinfo {author} {\bibfnamefont
  {S.}~\bibnamefont {Jia}},\ }\bibfield  {title} {\enquote {\bibinfo {title}
  {Spectroscopy of cesium Rydberg atoms in strong radio-frequency fields},}\
  }\href {http://dx.doi.org/10.1103/PhysRevA.94.023832} {\bibfield  {journal}
  {\bibinfo  {journal} {Phys. Rev. A}\ }\textbf {\bibinfo {volume} {94}},\
  \bibinfo {pages} {023832--1--7} (\bibinfo {year} {2016})}\BibitemShut
  {NoStop}%
\bibitem [{\citenamefont {Budker}\ and\ \citenamefont
  {Romalis}(2007)}]{Budker2007}%
  \BibitemOpen
  \bibfield  {author} {\bibinfo {author} {\bibfnamefont {D.}~\bibnamefont
  {Budker}}\ and\ \bibinfo {author} {\bibfnamefont {M.}~\bibnamefont
  {Romalis}},\ }\bibfield  {title} {\enquote {\bibinfo {title} {Optical
  magnetometry},}\ }\href {\doibase 10.1038/nphys566} {\bibfield  {journal}
  {\bibinfo  {journal} {Nature Physics}\ }\textbf {\bibinfo {volume} {3}},\
  \bibinfo {pages} {227--234} (\bibinfo {year} {2007})}\BibitemShut {NoStop}%
\bibitem [{\citenamefont {Daschner}\ \emph {et~al.}(2014)\citenamefont
  {Daschner}, \citenamefont {K\"ubler}, \citenamefont {L\"ow}, \citenamefont
  {Baur}, \citenamefont {Fr\"uhauf},\ and\ \citenamefont
  {Pfau}}]{Daschner2014}%
  \BibitemOpen
  \bibfield  {author} {\bibinfo {author} {\bibfnamefont {R.}~\bibnamefont
  {Daschner}}, \bibinfo {author} {\bibfnamefont {H.}~\bibnamefont {K\"ubler}},
  \bibinfo {author} {\bibfnamefont {R.}~\bibnamefont {L\"ow}}, \bibinfo
  {author} {\bibfnamefont {H.}~\bibnamefont {Baur}}, \bibinfo {author}
  {\bibfnamefont {N.}~\bibnamefont {Fr\"uhauf}}, \ and\ \bibinfo {author}
  {\bibfnamefont {T.}~\bibnamefont {Pfau}},\ }\bibfield  {title} {\enquote
  {\bibinfo {title} {Triple stack glass-to-glass anodic bonding for
  optogalvanic spectroscopy cells with electrical feedthroughs},}\ }\href
  {http://dx.doi.org/10.1063/1.4891534} {\bibfield  {journal} {\bibinfo
  {journal} {Appl. Phys. Lett.}\ }\textbf {\bibinfo {volume} {105}},\ \bibinfo
  {pages} {041107--1--4} (\bibinfo {year} {2014})}\BibitemShut {NoStop}%
\bibitem [{\citenamefont {Holloway}\ \emph
  {et~al.}(2014{\natexlab{b}})\citenamefont {Holloway}, \citenamefont {Gordon},
  \citenamefont {Jefferts}, \citenamefont {Schwarzkopf}, \citenamefont
  {Anderson}, \citenamefont {Miller}, \citenamefont {Thaicharoen},\ and\
  \citenamefont {Raithel}}]{Holloway2014}%
  \BibitemOpen
  \bibfield  {author} {\bibinfo {author} {\bibfnamefont {C.~L.}\ \bibnamefont
  {Holloway}}, \bibinfo {author} {\bibfnamefont {J.~A.}\ \bibnamefont {Gordon}},
  \bibinfo {author} {\bibfnamefont {S.}~\bibnamefont {Jefferts}}, \bibinfo
  {author} {\bibfnamefont {A.}~\bibnamefont {Schwarzkopf}}, \bibinfo {author}
  {\bibfnamefont {D.~A.}\ \bibnamefont {Anderson}}, \bibinfo {author}
  {\bibfnamefont {S.~A.}\ \bibnamefont {Miller}}, \bibinfo {author}
  {\bibfnamefont {N.}~\bibnamefont {Thaicharoen}}, \ and\ \bibinfo {author}
  {\bibfnamefont {G.}~\bibnamefont {Raithel}},\ }\bibfield  {title} {\enquote
  {\bibinfo {title} {Broadband Rydberg atom-based electric-field probe for
  SI-traceable, self-calibrated measurements},}\ }\href {\doibase
  10.1109/TAP.2014.2360208} {\bibfield  {journal} {\bibinfo  {journal} {IEEE
  Trans. Antennas Propag.}\ }\textbf {\bibinfo {volume} {62}},\ \bibinfo
  {pages} {6169--6182} (\bibinfo {year} {2014}{\natexlab{b}})}\BibitemShut
  {NoStop}%
\bibitem [{sco()}]{scopenote}%
  \BibitemOpen
  \href@noop {} {}\bibinfo {note} {The connection of an oscilloscope to the
  field plates to measure the plate voltage $V_C$ constitutes a change in
  boundary conditions that will result in a change in $t$. Also, since
  impedances are not well-defined, $V_C$ will generally not be the same as the
  voltage measured on the oscilloscope.}\BibitemShut {Stop}%
\bibitem [{\citenamefont {Anderson}\ \emph {et~al.}(2014)\citenamefont
  {Anderson}, \citenamefont {Schwarzkopf}, \citenamefont {Miller},
  \citenamefont {Thaicharoen}, \citenamefont {Raithel}, \citenamefont
  {Gordon},\ and\ \citenamefont {Holloway}}]{Anderson2014}%
  \BibitemOpen
  \bibfield  {author} {\bibinfo {author} {\bibfnamefont {D.~A.}\ \bibnamefont
  {Anderson}}, \bibinfo {author} {\bibfnamefont {A.}~\bibnamefont
  {Schwarzkopf}}, \bibinfo {author} {\bibfnamefont {S.~A.}\ \bibnamefont
  {Miller}}, \bibinfo {author} {\bibfnamefont {N.}~\bibnamefont {Thaicharoen}},
  \bibinfo {author} {\bibfnamefont {G.}~\bibnamefont {Raithel}}, \bibinfo
  {author} {\bibfnamefont {J.~A.}\ \bibnamefont {Gordon}}, \ and\ \bibinfo
  {author} {\bibfnamefont {C.~L.}\ \bibnamefont {Holloway}},\ }\bibfield
  {title} {\enquote {\bibinfo {title} {Two-photon microwave transitions and
  strong-field effects in a room-temperature Rydberg-atom gas},}\ }\href
  {\doibase 10.1103/PhysRevA.90.043419} {\bibfield  {journal} {\bibinfo
  {journal} {Phys. Rev. A}\ }\textbf {\bibinfo {volume} {90}},\ \bibinfo
  {pages} {043419--1--6} (\bibinfo {year} {2014})}\BibitemShut {NoStop}%
\bibitem [{\citenamefont {Anderson}\ \emph {et~al.}(2016)\citenamefont
  {Anderson}, \citenamefont {Miller}, \citenamefont {Raithel}, \citenamefont
  {Gordon}, \citenamefont {Butler},\ and\ \citenamefont
  {Holloway}}]{Anderson2016}%
  \BibitemOpen
  \bibfield  {author} {\bibinfo {author} {\bibfnamefont {D.~A.}\ \bibnamefont
  {Anderson}}, \bibinfo {author} {\bibfnamefont {S.~A.}\ \bibnamefont {Miller}},
  \bibinfo {author} {\bibfnamefont {G.}~\bibnamefont {Raithel}}, \bibinfo
  {author} {\bibfnamefont {J.~A.}\ \bibnamefont {Gordon}}, \bibinfo {author}
  {\bibfnamefont {M.~L.}\ \bibnamefont {Butler}}, \ and\ \bibinfo {author}
  {\bibfnamefont {C.~L.}\ \bibnamefont {Holloway}},\ }\bibfield  {title}
  {\enquote {\bibinfo {title} {Optical measurements of strong microwave fields
  with Rydberg atoms in a vapor cell},}\ }\href
  {http://dx.doi.org/10.1103/PhysRevApplied.5.034003} {\bibfield  {journal}
  {\bibinfo  {journal} {Phys. Rev. Appl.}\ }\textbf {\bibinfo {volume} {5}},\
  \bibinfo {pages} {034003--1--7} (\bibinfo {year} {2016})}\BibitemShut
  {NoStop}%
\bibitem [{\citenamefont {Miller}\ \emph {et~al.}(2016)\citenamefont {Miller},
  \citenamefont {Anderson},\ and\ \citenamefont {Raithel}}]{Miller2016}%
  \BibitemOpen
  \bibfield  {author} {\bibinfo {author} {\bibfnamefont {S.~A.}\ \bibnamefont
  {Miller}}, \bibinfo {author} {\bibfnamefont {D.~A.}\ \bibnamefont {Anderson}},
  \ and\ \bibinfo {author} {\bibfnamefont {G.}~\bibnamefont {Raithel}},\
  }\bibfield  {title} {\enquote {\bibinfo {title} {Radio-frequency-modulated
  Rydberg states in a vapor cell},}\ }\href
  {http://dx.doi.org/10.1088/1367-2630/18/5/053017} {\bibfield  {journal}
  {\bibinfo  {journal} {New J. Phys.}\ }\textbf {\bibinfo {volume} {18}},\
  \bibinfo {pages} {053017--1--8} (\bibinfo {year} {2016})}\BibitemShut
  {NoStop}%
\bibitem [{\citenamefont {Hill}\ \emph {et~al.}(1990)\citenamefont {Hill},
  \citenamefont {Kanda}, \citenamefont {Laren}, \citenamefont {Koepke},\ and\
  \citenamefont {Orr}}]{Hill1990}%
  \BibitemOpen
  \bibfield  {author} {\bibinfo {author} {\bibfnamefont {D.~A.}\ \bibnamefont
  {Hill}}, \bibinfo {author} {\bibfnamefont {M.}~\bibnamefont {Kanda}},
  \bibinfo {author} {\bibfnamefont {E.~B.}\ \bibnamefont {Laren}}, \bibinfo
  {author} {\bibfnamefont {G.~H.}\ \bibnamefont {Koepke}}, \ and\ \bibinfo
  {author} {\bibfnamefont {R.~D.}\ \bibnamefont {Orr}},\ }\bibfield  {title}
  {\enquote {\bibinfo {title} {Generating standard reference electromagnetic
  fields in the NIST anechoic chamber, 0.2 to 40 GHz},}\ }\href
  {https://archive.org/details/generatingstanda1335hill} {\bibfield  {journal}
  {\bibinfo  {journal} {NIST Technical Note 1335, National Institute of
  Standards and Technology, Boulder, CO, USA}\ } (\bibinfo {year}
  {1990})}\BibitemShut {NoStop}%
\bibitem [{\citenamefont {Matloubi}(1993)}]{Matloubi1993}%
  \BibitemOpen
  \bibfield  {author} {\bibinfo {author} {\bibfnamefont {K.}~\bibnamefont
  {Matloubi}},\ }\bibfield  {title} {\enquote {\bibinfo {title} {A broadband,
  isotropic, electric-field probe with tapered resistive dipoles},}\ }in\ \href
  {\doibase 10.1109/IMTC.1993.382655} {\emph {\bibinfo {booktitle}
  {Instrumentation and Measurement Technology Conference, 1993. IMTC/93.
  Conference Record., IEEE}}}\ (\bibinfo {year} {1993})\ pp.\ \bibinfo {pages}
  {183--184}\BibitemShut {NoStop}%
\bibitem [{\citenamefont {McGloin}\ \emph {et~al.}(2000)\citenamefont
  {McGloin}, \citenamefont {Dunn},\ and\ \citenamefont {Fulton}}]{McGloin2000}%
  \BibitemOpen
  \bibfield  {author} {\bibinfo {author} {\bibfnamefont {D.}~\bibnamefont
  {McGloin}}, \bibinfo {author} {\bibfnamefont {M. H.}\ \bibnamefont {Dunn}}, \
  and\ \bibinfo {author} {\bibfnamefont {D. J.}\ \bibnamefont {Fulton}},\
  }\bibfield  {title} {\enquote {\bibinfo {title} {Polarization effects in
  electromagnetically induced transparency},}\ }\href
  {http://dx.doi.org/10.1103/PhysRevA.62.053802} {\bibfield  {journal}
  {\bibinfo  {journal} {Phys. Rev. A}\ }\textbf {\bibinfo {volume} {62}},\
  \bibinfo {pages} {053802--1--6} (\bibinfo {year} {2000})}\BibitemShut
  {NoStop}%
\bibitem [{\citenamefont {Bao}\ \emph {et~al.}(2016)\citenamefont {Bao},
  \citenamefont {Zhang}, \citenamefont {Zhou}, \citenamefont {Zhang},
  \citenamefont {Zhao}, \citenamefont {Xiao},\ and\ \citenamefont
  {Jia}}]{Bao2016}%
  \BibitemOpen
  \bibfield  {author} {\bibinfo {author} {\bibfnamefont {S.}~\bibnamefont
  {Bao}}, \bibinfo {author} {\bibfnamefont {H.}~\bibnamefont {Zhang}}, \bibinfo
  {author} {\bibfnamefont {J.}~\bibnamefont {Zhou}}, \bibinfo {author}
  {\bibfnamefont {L.}~\bibnamefont {Zhang}}, \bibinfo {author} {\bibfnamefont
  {J.}~\bibnamefont {Zhao}}, \bibinfo {author} {\bibfnamefont {L.}~\bibnamefont
  {Xiao}}, \ and\ \bibinfo {author} {\bibfnamefont {S.}~\bibnamefont {Jia}},\
  }\bibfield  {title} {\enquote {\bibinfo {title} {Polarization spectra of
  Zeeman sublevels in Rydberg electromagnetically induced transparency},}\
  }\href {http://dx.doi.org/10.1103/PhysRevA.94.043822} {\bibfield  {journal}
  {\bibinfo  {journal} {Phys. Rev. A}\ }\textbf {\bibinfo {volume} {94}},\
  \bibinfo {pages} {043822--1--6} (\bibinfo {year} {2016})}\BibitemShut
  {NoStop}%
\bibitem [{\citenamefont {Zhang}\ \emph {et~al.}(2017)\citenamefont {Zhang},
  \citenamefont {Bao}, \citenamefont {Zhang},\ and\ \citenamefont
  {Raithel}}]{Zhang2017}%
  \BibitemOpen
  \bibfield  {author} {\bibinfo {author} {\bibfnamefont {L.}~\bibnamefont
  {Zhang}}, \bibinfo {author} {\bibfnamefont {S.}~\bibnamefont {Bao}}, \bibinfo
  {author} {\bibfnamefont {H.}~\bibnamefont {Zhang}}, \ and\ \bibinfo {author}
  {\bibfnamefont {G.}~\bibnamefont {Raithel}},\ }\href@noop {} {} (\bibinfo
  {year} {2017}),\ \Eprint {http://arxiv.org/abs/arXiv:1702.04842}
  {arXiv:1702.04842} \BibitemShut {NoStop}%
\end{thebibliography}

%

\end{document}